\documentclass[aps,prd,preprint,showpacs,showkeys]{revtex4}

\usepackage{latexsym}

\begin{document}


\title{Note on an action for a particle in the Ho\v{r}ava-Lifshitz Gravity}

\author{Myungseok Eune}
\email[]{younms@sejong.ac.kr}
\affiliation{Institute of Fundamental Physics, Sejong University, Seoul 143-747, Korea}

\author{Wontae Kim}
\email[]{wtkim@sogang.ac.kr}
\affiliation{Department of Physics,
  Sogang University, Seoul, 121-742, Korea}

\date{\today}

\begin{abstract}
  We reconsider a recently proposed action for a free particle which
  is compatible with Ho\v{r}ava-Lifshitz gravity, and then obtain the
  subluminal and the superluminal limits without gauge ambiguity in
  terms of Hamiltonian formulation.
\end{abstract}



\maketitle


Recently, Ho\v{r}ava has introduced a (power-counting) renormalizable
theory of gravity called Ho\v{r}ava-Lifshitz (HL) gravity, where the
scaling dimensions of space and time are different
\cite{Horava:2008jf,Horava:2008ih,Horava:2009uw}. Subsequently, there
have been extensive studies for the HL gravity
\cite{Cai:2009pe,Myung:2009va,Mann:2009yx,Cai:2009ar,Colgain:2009fe,Ghodsi:2009zi,Kiritsis:2009vz,Majhi:2009xh,Romero:2009qs,Takahashi:2009wc,Brandenberger:2009yt,Mukohyama:2009zs,Alexander:2009uu,Setare:2010wt,Jamil:2010di,mosaffa},
in particular, an interesting action describing a free particle
reflecting Foliation Preserving Diffeomorphism (FPD) has been proposed
and its geodesic motion of the particle has been studied in
Ref.~\cite{mosaffa}.  However, there appear the two types of gauge
fixing conditions simultaneously: one is for the auxiliary variable
$e(\tau)$ and the other is for reparametrization symmetry. Eventually,
it means that the system is over-constrained so that physical
interpretations may depend on the gauge fixing condition.  So, in this
paper, we would like to resolve this problem using the Hamiltonian
formulation by treating constraints appropriately~\cite{dirac,ht}.
The reparametrization invariance indicates that the constraint system
is a first class. By fixing a gauge consistently, the constraint
becomes fully second class. So, it does not need any additional gauge
fixing condition.

Now, assuming the line element, $ds^2 = -N^2 c^2 dt^2 + g_{ij} du^i
du^j$, where $ du^i = dx^i + N^i dt$, ($i = 1, 2, 3$) and $N$ and
$N^i$ are the usual lapse and shift functions, the action for the free
particle which reflects the FPD is given by~\cite{mosaffa}
\begin{equation}
  \label{action}
  S = \frac12 \int d\tau \left[ \frac{1}{e} \left( g_{ij} \dot{u}^i \dot{u}^j -
    c^2 N^2 \dot{t}^2 \right) + \frac{eM^2}{N^2\dot{t}^2} \left(
    g_{ij} \dot{u}^i \dot{u}^j - \frac{m^2}{M^2} c^2 N^2 \dot{t}^2
  \right) \right],
\end{equation}
where $\tau$ is the time to parametrize the world line of the
particle, $e(\tau)$ is the worldline einbein, and the overdot denotes
the derivative with respect to $\tau$. The action~(\ref{action}) is
invariant under the reparametrization given by
\begin{equation}
  \label{symmetry}
  \tau \to \tau'(\tau), \quad e(\tau) \to e'(\tau') = e(\tau)
  \frac{d\tau}{d\tau'}, \quad (x^i, t) \to (x^i, t). 
\end{equation}
Note that for the limit of $c \to 0$, the action (\ref{action})
becomes the UV action, while for the limit of $M \to 0$, it is just
the IR action as explicitly presented in Ref.~\cite{mosaffa}.

Let us simply consider the flat case of $N=1$ and $g_{ij} =
\delta_{ij}$. Then, the Lagrangian from Eq.~(\ref{action}) can be
written as
\begin{equation}
  \label{L}
  L = \frac12 \left[ \frac{1}{e} \left( \dot{u}^i \dot{u}_i -
    c^2 \dot{t}^2 \right) + \frac{eM^2}{\dot{t}^2} \left(
    \dot{u}^i \dot{u}_i - \frac{m^2}{M^2} c^2   \dot{t}^2
  \right) \right].
\end{equation}
The conjugate momenta of $u^i$, $e$, and $t$ are obtained as
\begin{eqnarray}
  p_i &=& \left( \frac{1}{e} + \frac{eM^2}{  \dot{t}^2} \right)
  \dot{u}_i, \label{p_i} \\
  p_e &=& 0, \label{p_e} \\
  p_t &=& - \frac{eM^2}{  \dot{t}^3} \dot{u}^i \dot{u}_i -
  e^{-1} c^2 \dot{t}, \label{p_t:def}
\end{eqnarray}
which yields one primary constraint,
\begin{equation}
  \label{Omega_1}
  \Omega_1 = p_e \approx 0.
\end{equation}
Combining Eq.~(\ref{p_i}) and Eq.~(\ref{p_t:def}), we can get the
momentum--velocity relation,
\begin{equation}
  \label{p_t}
  p_t = - \frac{eM^2}{\dot{t}^3} \left( \frac{1}{e} +
    \frac{eM^2}{\dot{t}^2} \right)^{-2} p^i p_i - e^{-1} c^2  \dot{t}.
\end{equation}
However, it is not easy to get the inverse relation so that we
formally write $\dot{t} \equiv \eta (e,p_i, p_t)$ and $\dot{u} \equiv
\xi(e,p_i,p_t)$.  Then, the primary Hamiltonian can be written as
\begin{equation}
  \label{H_p}
  H_p = H_c(e,p_i,p_t) + \lambda \Omega_1,
\end{equation}
where $\lambda$ is the Lagrange multiplier and the canonical Hamiltonian
$H_c$ is given by
\begin{equation}
  \label{H_c}
  H_c = \frac12 \left( \frac{1}{e} + \frac{eM^2}{\eta^2}
  \right)^{-1} p^i p_i + p_t \eta + \frac12 ( e^{-1} c^2  
  \eta^2 + e m^2 c^2).
\end{equation}
To discuss the stability of time evolution of the primary constraint
(\ref{Omega_1}), let us define Poisson brackets between the variables
as followings,
\begin{equation}
  \label{PB:def}
   \{u^i, p_j \} = \delta^i_j, \quad \{e, p_e\} = 1, \quad \{t, p_t\} =
  1, \quad \mathrm{others} = 0.
\end{equation}
Then, using the convenient relation of $\partial \eta / \partial e = e^{-1}
\eta$ from Eq.~(\ref{p_t}), the time evolution of the primary
constraint gives the secondary constraint as
\begin{equation}
  \label{Omega_2}
  \dot{\Omega}_1 = \{\Omega_1, H_p \} = - \frac{1}{e} H_c (e,
  p_i, p_t)  = \frac{1}{e} \Omega_2 \approx 0.
\end{equation}
It happens that there is no more secondary constraint because the time
evolution of $\Omega_2$ automatically vanishes, $\dot{\Omega}_2 =
\lambda e^{-1} \Omega_2 \approx 0$.  Moreover, the primary and the
secondary constraint are the first class which is necessarily the
implementation of the reparametrization symmetry.

Now, one can fix the gauge to make the system into second class so that take the
gauge fixing condition as
\begin{equation}
  \label{Omega_3}
  \Omega_3 \equiv t - a\tau \approx 0,
\end{equation}
with a gauge parameter $a$.  It means that $\dot{t} = \eta =a$.
Furthermore, the careful time evolution of the gauge fixing condition
(\ref{Omega_3}),
\begin{equation}
  \label{Omega_3:dot}
  \dot{\Omega}_3 = \{ \Omega_3, H_p \} + \partial
\Omega_3/ \partial \tau = \left( \frac{\partial \eta}{\partial p_t}
\right) \Omega_4 + \dot{t} - a \approx 0,
\end{equation}
yields additional constraint,
\begin{equation}
  \label{Omega_4}
  \Omega_4 \equiv p_t + \frac{eM^2}{a^3} \left( \frac{1}{e} +
    \frac{eM^2}{a^2} \right)^{-2} p^i p_i + e^{-1} c^2   a
  \approx 0.
\end{equation}
Requiring that the time evolution of $\Omega_4$ vanishes, 
we can fix the Lagrangian multiplier as $\lambda =0$.  Thus, we can obtain
the four constraints which give the well-defined Dirac brackets.
Applying the gauge fixing condition (\ref{Omega_3}), the canonical
Hamiltonian~(\ref{H_c}) can be rewritten as
\begin{equation}
  \label{H_c:gauge}
  H_c = \frac12 \left( \frac{1}{e} + \frac{eM^2}{a^2}
  \right)^{-1} p^i p_i + a p_t + \frac12 ( e^{-1} c^2  
  a^2 + e m^2 c^2).
\end{equation}

After some calculations, one can now obtain constraint algebra $C_{ab}
\equiv \{ \Omega_a, \Omega_b \} $ as
\begin{equation}
  \label{C}
  C = \left(
  \begin{array}{cccc}
    0 & 0 & 0 & - \Delta \\
    0 & 0 & -a & 0 \\
    0 & a & 0 & 1 \\
    \Delta & 0 & -1 & 0
  \end{array} \right),
\end{equation}
where
\begin{equation}
  \label{Delta:def}
  \Delta \equiv  \frac{\partial \Omega_4}{\partial e} =
  - \frac{M^2}{a^3} \left( \frac{1}{e} + \frac{eM^2}{a^2}
  \right)^{-3} \left( - \frac{3}{e} + \frac{eM^2}{a^2} \right) p^i
  p_i - e^{-2} c^2   a.
\end{equation}
Since $\det C = a^2 \Delta^2 \ne 0$, the constraint algebra is fully
second class, which implies that we can not fix additional gauge
anymore, for instance, $e(\tau)=1/m$.  Therefore, Dirac brackets given
by $\{u, v \}_{\rm DB} \equiv \{u, v \} - \sum_{a,b} \, \{u, \Omega_a\}
C^{-1}_{ab} \{\Omega_b, v\}$, are well-defined so that nontrivial
brackets are exhibited as
\begin{eqnarray}
  \{e, t\}_{\rm DB} &=&  \frac{1}{\Delta}, \label{DB:et} \\
  \{e, u^i\}_{\rm DB} &=& \frac{2eM^2}{\Delta   a^2} \left(
    \frac{1}{e} + \frac{eM^2}{  a^2} \right)^{-2} p^i, \label{DB:eu}
  \\
  \{u^i, p_j\}_{\rm DB} &=& \delta^i_j, \label{DB:up} \\
  \{u^i, p_t\}_{\rm DB} &=& - \frac{1}{a} \left(
    \frac{1}{e} + \frac{eM^2}{  a^2} \right)^{-1} p^i. \label{DB:up_t}
\end{eqnarray}

Finally, using the fact that $\dot{u}^i = du^i/d\tau = \eta\, du^i/dt = a v^i$ with $v^i
\equiv du^i/dt$, Eq.~(\ref{Omega_2}) leads to
\begin{equation}
  \label{eom}
  e^2 \left(\frac{M^2 v^2}{c^2} - m^2 \right) = a^2
  \left(\frac{v^2}{c^2} - 1 \right).
\end{equation}
For $M<m$, the allowed range of velocity is $v<c$ (subluminal) or
$v>\frac{m}{M}c$ (superluminal). In the subluminal regime $v<c$ with
$M \ll m$, the action goes to the IR action. For $M > m$, the allowed
range of velocity is $v<\frac{m}{M} c$ (subluminal) or $v>c$
(superluminal). The superluminal case of them corresponds to the UV
action along with $c\to 0$. For the luminal case $v=c$, the masses are
zero, \textit{i.e.,} $M=m=0$. Note that all these arguments are
gauge-independent.

In conclusion, we have studied the recently proposed action for a free
particle where the scaling dimensions of space and time are different.
Note that in the IR region of $M \approx 0$ which recovers the
relativistic particle action, the subluminal speed of the particle can
be well-defined, however, there exists superluminal region with the
infinite speed in spite of the massive particle. On the other hand, in
the UV region defined by $c \approx 0$, the speed of the particle is
expectedly unlimited.

\begin{acknowledgments}
  We would like to thank E.\ Son for exciting discussions.  WT Kim was
  supported by the Special Research Grant of Sogang University,
  200911044.  M.\ Eune was supported by the National Research
  Foundation of Korea Grant funded by the Korean Government
  [NRF-2009-351-C00109].
\end{acknowledgments}



\end{document}